\documentclass[11pt,a4paper]{article}
\usepackage{jheppub}

\usepackage{latexsym,amsmath,amsfonts,amsthm,amssymb,bm,bbm} 
\usepackage{braket} 
\usepackage{graphicx} 
\usepackage[usenames, dvipsnames, svgnames]{xcolor}
\usepackage{tikz}
\usetikzlibrary{intersections, calc, patterns}

\usepackage{jheppub} 

\newcommand{\Tr}{\mathrm{Tr}} 

\def\CS{{\cal S}}

\def\BH{\mathbb{H}}

\def\BS{\mathbb{S}}
\def\BZ{\mathbb{Z}}

\title{A Holographic Proof of R\'enyi Entropic Inequalities}

\author[a,b]{Yuki Nakaguchi}
\author[b]{and Tatsuma Nishioka}

\affiliation[a]{Kavli Institute for the Physics and Mathematics of the Universe,
The University of Tokyo,\\
5-1-5 Kashiwa-no-Ha, Kashiwa City, Chiba 277-8568, Japan}
\affiliation[b]{Department of Physics, Faculty of Science,
The University of Tokyo,\\
Bunkyo-ku, Tokyo 113-0033, Japan}

\emailAdd{yuuki.nakaguchi@ipmu.jp}
\emailAdd{nishioka@hep-th.phys.s.u-tokyo.ac.jp}

\abstract{
We prove R\'enyi entropic inequalities in a holographic setup
based on the recent proposal for the holographic formula of R\'enyi entropies
when the bulk is stable against any perturbation.
Regarding the R\'enyi parameter as an inverse temperature, we reformulate the entropies in analogy with statistical mechanics, which provides us a concise interpretation of the inequalities as the positivities of entropy, energy and heat capacity.
This analogy also makes clear a thermodynamic structure in deriving the holographic
formula. As a by-product of the proof we obtain a holographic formula to calculate 
the quantum fluctuation of the modular Hamiltonian.
A few examples of the capacity of entanglement are examined in detail.
}

\preprint{IPMU-16-0090, UT-16-26}

\begin{document}

\maketitle

\section{Introduction} 
A key concept in modern quantum gravity theory is holography that opened the door to the non-perturbative definition as the dual quantum theory in one lower dimensions.
A considerable number of dictionaries have been composed to translate physical quantities in one theory to the other. The holographic duality remains as mysterious as quantum gravity,
though, especially on how the bulk spacetime information is encoded in the boundary
quantum field theory.
There have been a huge amount of attempts to probe the bulk structure via holography, of which one of the most important breakthroughs is the
holographic formula of entanglement entropy \cite{Ryu:2006bv} associating a unit area per four times the Planck length of a codimension-two bulk surface with one bit of information for a given region in the boundary field theory.
In fact, the formula is a realization of the original idea of the holographic principle \cite{Susskind:1994vu,'tHooft:1993gx}
that states in quantum gravity theory, the degrees of freedom live not in volumes but in areas.
Overviews on the recent developments of the holographic entanglement entropy are available in reviews e.g. \cite{Takayanagi:2012kg,Harlow:2014yka}.

In quantum theories, entanglement entropy $S_A$
of a state subspace $\mathcal{H}_A$ is defined as
the von Neumann entropy
$S_\text{vN}[\rho] \equiv -\Tr[\rho\log\rho]$
of the reduced matrix $\rho_A=\Tr_{\bar{A}}[\rho_\text{total}]$ as\footnote{
Throughout this paper, we always normalize a density matrix as $\Tr[\rho]=1$.}
\begin{align}
  S_A &\equiv -\Tr[\rho_A\log\rho_A]\,. \label{EE}
\end{align}
It measures how much quantum information of the degrees of freedom in $\mathcal{H}_A$ is entangled with the outer degrees of freedom,
namely, how much quantum information will be lost for the subspace $\mathcal{H}_A$ if the outer subspace is ignored.
In quantum field theories,
entanglement entropy is defined for a space region $A$ on a time slice, assuming that we can construct a state space $\mathcal{H}_A$ representing degrees of freedom on the region $A$ by some appropriate procedures.
The total state is often taken as the vacuum $\rho_\text{total}=\ket{0}\bra{0}$
for simplicity.

Entanglement entropy has many mathematical properties, among which the most important one is an inequality called the strong sub-additivity \cite{Lieb:1973cp}
\begin{align}
  S_{AC} +S_{BC} &\ge S_{C}+S_{ABC}\,, \label{SSA}
\end{align}
showing a kind of concavity of the entropy.
The sub-additivity
\begin{align}
  S_A+S_B\ge S_{AB} \label{SA}
\end{align}
follows by taking $C$ as $\emptyset$.
As a field application, the strong sub-additivity is utilized for constructing $c$-functions, monotonically decreasing functions along RG flows, such as the entropic $c$-function in two dimensions \cite{Casini:2006es} and the $F$-function in three dimensions \cite{Casini:2012ei}.

One of the novel aspects of the holographic entanglement entropy formula is the simplicity of proving the strong sub-additivity
\eqref{SSA} \cite{Headrick:2007km,Wall:2012uf,Headrick:2013zda}.
The proof only relies on the geometric properties of a codimension-two surface in the bulk, and suggests a profound way of the emergence of the bulk spacetime as it translates a quantum mechanical constraint to a purely geometric one.
More extensive studies of the inequalities satisfied by the holographic formula were carried out in \cite{Hayden:2011ag,Bao:2015bfa} to classify the characteristics of the geometry which has a field theory dual.

Recently,
the holographic formula was proposed \cite{Dong:2016fnf} for the entanglement R\'enyi entropy $S_n[\rho]$ which is a one-parameter generalization of the von Neumann entropy defined with a non-negative real number $n$ as
\begin{align}
  S_n[\rho] &\equiv -\frac{1}{n-1} \log\Tr[\rho^n] \label{renyi}\,.
\end{align}
It reduces to the von Neumann entropy when $n=1$, $S_1[\rho] = S_\mathrm{vN}[\rho]$.
The derivation of the holographic formula by \cite{Dong:2016fnf} is based on so-called the Lewkowycz-Maldacena prescription \cite{Lewkowycz:2013nqa} employed to derive the holographic entanglement entropy where the replica $\BZ_n$ symmetry is assumed in the bulk geometry.\footnote{Earlier works on the holographic R{\'e}nyi entropies include \cite{Fursaev:2006ih,Headrick:2010zt,Hung:2011nu}.}
We will review the derivation in section \ref{ss:HERformula} so as to fix our notations and for later use.

Then it is natural to think about how mathematical properties of the R\'enyi entropy are transcribed to the bulk side in a geometric language.
It is known that the R\'enyi entropy is not strongly sub-additive, but
it satisfies inequalities involving the derivative with respect to $n$ \cite{zyczkowski2003renyi,Beck}\footnote{The finite version of these inequalities, such as $S_n\ge S_m$ and $\frac{n-1}{n}S_n \ge \frac{m-1}{m}S_m$ for $n \le m$, are true, even if the $n$ derivatives are ill-defined because of some discontinuity.}
\begin{align}
  \partial_n S_n &\le 0 \label{inequality1}\,, \\
  \partial_n \left(\frac{n-1}{n}S_n\right) &\ge 0 \label{inequality2} \,, \\
  \partial_n \left((n-1)S_n\right) &\ge 0 \label{inequality3} \,, \\
  \partial_n^2 \left((n-1)S_n\right) &\le 0 \label{inequality4} \,.
\end{align}
These inequalities are originally proved for the classical R\'enyi entropy
$
  S_n[p_i] \equiv -\frac{1}{n-1} \sum_i p_i^n
$
of a probability distribution $p_i$,
but are still true for the quantum R\'enyi entropy \eqref{renyi}.
The proof for a quantum case immediately follows by diagonalizing the density matrix $\rho$ as
$U \rho\, U^\dagger = \text{diag}(p_1,p_2,\dots)$ with a unitary matrix $U$.
The first inequality \eqref{inequality1} implies the positivity of the R\'enyi entropy $S_n\ge 0$ as $S_\infty=\text{min}_i(-\log p_i)\ge0$. 

The aim of this paper is to prove these inequalities by the holographic formula of the R\'enyi entropy.
Before proceeding to the proof, we rewrite the inequalities in more concise forms that manifest their meanings as the positivities of energy, entropy and heat capacity in analogy to statistical mechanics.
It also clarifies that not all of \eqref{inequality1}-\eqref{inequality4} are independent, but the two inequalities \eqref{inequality2} and \eqref{inequality4} are essential.
\eqref{inequality2} turns out to be simple to prove as it stands for the positivity of the area of a codimension-two surface in the bulk, while the proof of \eqref{inequality4} is more intricate.
In view of statistical mechanics, \eqref{inequality4} implies the positivity of the heat capacity and encodes the unitarity of quantum mechanical system.
Our proof of \eqref{inequality4} in the bulk is differential geometric in its nature and turns out to relate it to the stability of the spacetime on which the holographic formula is supposed to be applied.
Therefore, our proof serves as a nontrivial consistency check for the holographic formula, and moreover reveals a direct connection between the unitarity and the stability in the boundary and bulk theories, respectively.
In due course of the proof, we also obtain a holographic formula for calculating the quantum fluctuation of the modular Hamiltonian.

Our proof is heavily based on the stability of the bulk geometry. We admit that the bulk stability is a nontrivial assumption whose justification is even challenging. For instance, Euclidean gravity actions are known to be indefinite against metric perturbations \cite{Gibbons:1978ac,Christensen:1979iy}. We are not aware of any compelling argument to support the assumption, but in view of holographic duality we believe that stable quantum states should have stable bulk duals. We will not touch on this subject anymore in this paper until section \ref{ss:Discussion}.

The organization of this paper is as follows.
In section \ref{ss:Stat} we reformulate the R\'enyi entropy and its inequalities in a way analogous to statistical mechanics
and introduce a notion of the heat capacity of entanglement.
The holographic formula of the R\'enyi entropy is reviewed in section \ref{ss:HERformula} with emphasis on the analogy to statistical mechanics.
In section \ref{ss:Proof}, we prove the R\'enyi entropic inequalities from the holographic point of view. 
The capacity of entanglement is exemplified in various systems in section \ref{ss:Capacity}.
Finally section \ref{ss:Discussion} is devoted to discussions on our results and future directions.
Appendix \ref{ap:Graviton} deals with an alternative method of computing the capacity of entanglement in the holographic setup and discusses a delicate issue arising from boundary terms.
A possible counterpart of the strong sub-additivity for the R\'enyi entropy is elaborated in appendix \ref{ap:SSA_RE}.

\section{Analogy to statistical mechanics}\label{ss:Stat}
The R\'enyi entropy can be recasted as a thermal entropy when the region $A$ is a ball in CFT$_d$ as the replica manifold $\mathcal{M}_n$ is conformally equivalent to a thermal hyperbolic space $\mathbb{S}^1\times \mathbb{H}^{d-1}$ with an inverse temperature $\beta=2\pi n$ \cite{Hung:2011nu,Casini:2011kv}. 
In that situation, the inequalities \eqref{inequality2} and \eqref{inequality4} reduce to the non-negativity of the thermal entropy and the heat capacity, and the others immediately follow from these two.
A formal similarity between the R\'enyi entropy and a thermal entropy is also pointed out in \cite{Dong:2016fnf}.

In this section, inspired by these observations,
we will formulate the complete analogy between the R\'enyi entropy and statistical mechanics
valid for any quantum system.
Moreover, the following discussions apply not only to reduced density matrices $\rho_A = \Tr_{\bar{A}}[\rho_\text{tot}]$, but also to a general density matrix $\rho$.

\subsection{Partition function $Z$ and the escort density matrix $\rho_n$}

In the calculation of the R\'enyi entropy $S_n=-\frac{1}{n-1}\log\Tr[\rho^n]$ \eqref{renyi},
we can regard the trace $Z(n)\equiv\Tr[\rho^n]$ as a thermal partition function
\begin{align}
  Z(\beta)=\Tr[e^{-\beta H}] \,,
\end{align}
with\footnote{
If you feel uneasy about the mismatch of their physical dimensions,
you may define them instead as
\begin{align}
  \beta E_0 &= n \,,\\
  H / E_0 &= - \log \rho \,,
\end{align}
with any constant $E_0$ of the dimension of energy.
In the following discussions we take a unit $E_0 = 1$.
Another choice $E_0=1/2\pi$ is also common in literatures.
}
the inverse temperature $\beta$ and the Hamiltonian $H$
\begin{align}
  \beta &= n \,,\\
  H &= - \log \rho \,.
\end{align}
The latter is called the entanglement Hamiltonian or modular Hamiltonian. 
Its eigenvalues $\epsilon_i$ are called the entanglement spectrum,
and are non-negative $\epsilon_i\ge 0$
as the eigenvalues $p_i=e^{-\epsilon_i}$ of $\rho$ satisfies $0\le p_i \le 1$.
In calculating the partition function $Z$,
we can regard the state as a density matrix given by the normalized $n$-th power of $\rho$
\begin{align}
  \rho_n \equiv \frac{\rho^n}{\Tr[\rho^n]}\,.
\end{align}
In the area of chaotic systems, 
the probability distribution of the classical version
$ P_i^{(n)} \equiv p_i^n/\sum_i p_i^n$
is called the escort distribution \cite{Beck},
and we will accordingly call $\rho_n$ the escort density matrix.


Let us push forward this analogy to statistical mechanics.
The free energy $F=F(n)$ and the total energy $E=E(n)$ related to the density matrix $\rho$ are defined as
\begin{align}
  F &\equiv -\frac{1}{n} \log \Tr[\rho^n] \,, \label{FreeEnergy}\\
  E &\equiv -\frac{\partial}{\partial n} \log \Tr[\rho^n] = \braket{ H}_n \label{energy}\,,
\end{align}
where $\braket{\cdot}_n$ stands for the expectation values with respect to the escort density matrix $\rho_n$,
\begin{align}
  \braket{X}_n \equiv \Tr[\rho_n X]
  = \frac{\Tr[\rho^nX]}{\Tr[\rho^n]} \,.
\end{align}
In what follows, we will make use of this notation when available.

\subsection{Improved R\'enyi entropy $\tilde{S}_n$}
What quantity should correspond to the thermal entropy in this analogy to statistical mechanics?
The answer is not the R\'enyi entropy $S_n[\rho]$,
but a more involved function:
\begin{align}
  \tilde{S}_n[\rho] &\equiv
  n^2 \partial_n \left(
    \frac{n-1}{n}S_n
  \right) \label{improved1} \,,\\
  &= \left(1 - n\partial_n\right) \log \Tr[\rho^n] \label{improved2}\,,\\
  &= \partial_{1/n}\left(
    \frac{1}{n} \log \Tr[\rho^n]
  \right) \label{improved3}\,.
\end{align}
Let us call this $\tilde{S}_n[\rho]$ as the improved R\'enyi entropy.
In fact, the equation \eqref{improved2} or \eqref{improved3}
leads to the formulae of the entropy together with \eqref{FreeEnergy} and \eqref{energy},
\begin{align}
\begin{aligned}
  \tilde S & = n(E-F)\,, \\
  &= -\frac{\partial F}{\partial T}
  \,,
\end{aligned}
\end{align}
where $T \equiv 1/n$ and we omit the subscript of $\tilde S_n$ to stress the correspondence to statistical mechanics.
One can also show that the improved R\'enyi entropy is nothing but the von Neumann entropy
 of the escort density matrix $\rho_n$, that is,
\begin{align}
  \tilde{S}_n[\rho] =
  S_\mathrm{vN}[\rho^n/\Tr[\rho^n]]\,.
\end{align}
The improved R\'enyi entropy $\tilde{S}_n$ is another generalization of the von Neumann entropy $S_\text{vN}$ as it also reduces to the entanglement entropy $\tilde{S}_1[\rho]=S_\mathrm{vN}[\rho]$
in the limit $n\to1$.

An equivalent relation to \eqref{improved1}
\begin{align}
  (n-1)^2\partial_n S_n
    &= \tilde{S}_n - E \label{delS1}\,,
\end{align}
yields a useful formula for calculating $\partial_n S_n$ in terms of $F$
\begin{align}
  \partial_n S_n
    &= T^2\frac{F(1) - F(T) - ( 1 - T )\partial_T F}{(1-T)^2}\,, \label{delS2}
\end{align}
where we used the relations $E=F+T\tilde{S}$, $\tilde{S}=-\partial_T F$ and $F(1)=0$.

\subsection{Capacity of entanglement $C(n)$}
Now that we have defined thermodynamic state functions consisting of the first derivative of the free energy such as the total energy $E=\partial_n(nF)$ and the thermal entropy $\tilde{S}=-\partial_T F$,
we proceed to implement the heat capacity $C=C(n)$ including the second derivative,
\begin{align}
  C \equiv \frac{\partial E}{\partial T}
  = T\frac{\partial \tilde{S}}{\partial T}
  = -T\frac{\partial^2F}{\partial T^2} \,.
\end{align}
It was originally introduced to characterize topologically ordered states by \cite{PhysRevLett.105.080501} and named capacity of entanglement.
The capacity of entanglement has not attracted much attention so far despite its importance and simplicity as we will see below.

One can show the non-negativity $C \ge 0$ as in the same way as statistical mechanics,
\begin{align}
\begin{aligned}
  C(n) =n^2\frac{\partial^2}{\partial n^2}\log Z(n)
  &= n^2(\braket{H^2}_n-\braket{H}_n^2) \,,\\
  &= n^2\braket{(H-\braket{H}_n)^2}_n \label{positivity_of_c} 
   \,.
\end{aligned}
\end{align}
It follows that the capacity measures the quantum fluctuation of the modular Hamiltonian $H=-\log\rho$, and in particular
$C(1)=\braket{H^2}-\braket{H}^2$ gives the quantum fluctuation with respect to the original state $\rho$.

\subsection{R\'enyi entropic inequalities from the viewpoint of the analogy}
Having established the analogy to statistical mechanics,
we rewrite the R\'enyi entropic inequalities in the thermodynamic representation.
The second \eqref{inequality2}, third \eqref{inequality3} and forth \eqref{inequality4} inequalities turn out to be the non-negativity of the improved R\'enyi entropy $\tilde{S}$ \eqref{improved1},
the total energy $E$ \eqref{energy}
and the entanglement heat capacity $C$, respectively
\begin{align}
  \tilde{S}  &\ge 0 \label{inequality2_thermo}\,,\\
  E &\ge 0 \,, \label{inequality3_thermo}\\
  C &\ge 0\,. \label{inequality4_thermo}
\end{align}
The non-negativity of $\tilde{S}$ and $E = \braket{H}_n$ immediately follows from the relations
$  \tilde{S}_n[\rho] =\braket{-\log\rho_n}_n= S_\mathrm{vN}[\rho_n]$ and the definition $H=-\log\rho$ of the modular hamiltonian.
The last inequality $C\ge 0$ has already been proved by \eqref{positivity_of_c}.
Note that the condition $C\ge0$ is equivalent to 
\begin{align}
  \partial_n\tilde{S}_n \le 0\,,
\end{align}
because of $C = T\partial_T\tilde{S} = -n\partial_n \tilde{S}$.

The first inequality $\partial_nS_n\le 0$  \eqref{inequality1} can be derived from the forth inequality \eqref{inequality4} as shown in \cite{Hung:2011nu}.
Indeed, the forth inequality
$C = -T \partial_T^2 F \ge 0$
is equivalent to the concavity
of the free energy $F$, and \eqref{delS2} is clearly non-positive
as $f(x) \le f(a) + (x-a)f'(a)$ for any concave function $f(x)$.
An alternative way to show this inequality uses the non-negativity of the relative entropy
$  S[\rho|\sigma]\equiv \Tr[\rho(\log\rho-\log\sigma)]\ge 0$ for \eqref{delS1}
\begin{align}
\begin{aligned}
  (n-1)^2\partial_n S_n &= \tilde{S}_n - E \,,\\
    &=-\braket{\log\rho_n-\log \rho}_n \,,\\
    &=-S[\rho_n|\rho] \le 0\,.
\end{aligned}
\end{align}

\section{Holographic formula of the R\'enyi entropy}\label{ss:HERformula}
We review the holographic formula for the R\'enyi entropy and its derivation proposed by \cite{Dong:2016fnf}, with some clarifications on the thermodynamic interpretation developed in the previous section.
It resembles to the Ryu-Takayanagi formula, but is more intricate as
the entropy is given by the area of a cosmic brane with a tension depending on the parameter $n$, which is extremized in the backreacted geometry.
The derivation of the formula still proceeds along with the Lewkowycz-Maldacena prescription \cite{Lewkowycz:2013nqa}.

\subsection{The area prescription}
The holographic formula for the R\'enyi entropy \cite{Dong:2016fnf} states that
the improved R\'enyi entropy $\tilde{S}_n$ of a region $A$ in QFT$_d$
is given by
the area $\mathcal{A}$ of a codimension-two surface $C_A^{(n)}$ in an asymptotically AdS$_{d+1}$ space as
\begin{align}
  \tilde{S}_n =
  \frac{\mathcal{A}}{4G_N}
  \bigg|
    _{\delta I = 0,\,\partial C_A^{(n)} = \partial A}\,, \label{HRE}
\end{align}
where the surface $C_A^{(n)}$ is anchored on $\partial A$ on the asymptotic boundary of the bulk spacetime.
Unlike the Ryu-Takayanagi formula, 
the surface $C_A^{(n)}$ is to be fixed by minimizing an $n$-dependent Euclidean action
$I = I_\text{bulk} + I_\text{brane}$.
Here  $I_\text{bulk}$ is
the original bulk action in the dual gravity theory consisting of the Einstein-Hilbert action, the cosmological constant term and matter terms
\begin{align}
  I_\text{bulk}[G_{\mu\nu}(X),\psi(X)] = I_\text{EH}[G_{\mu\nu}(X)] + I_\Lambda[G_{\mu\nu}(X)] + I_\text{matters}[G_{\mu\nu}(X),\psi(X)]\,,
\end{align}
where  $G_{\mu\nu}(X)$ the bulk metric,
$\psi(X)$ matter fields, and $X^\mu$ ($\mu=0, \dots, d$) is the bulk coordinate.
If we extremize the codimension-two surface with this bulk action, we end up with the Ryu-Takayanagi surface for the holographic entanglement entropy.
A new ingredient of the prescription for the R{\'e}nyi entropy is a cosmic brane action $I_\text{brane}$ of $C_A^{(n)}$,
\begin{align}
  I_\text{brane}[G_{\mu\nu}(X),X^\mu(y)]
  = T_n\,\mathcal{A}[G_{\mu\nu}(X),X^\mu(y)]\,,
\end{align}
which is just the product of 
a brane tension $T_n$ given by
\begin{align}
  T_n = \frac{1}{4G_N}\frac{n-1}{n}\,, \label{tension}
\end{align}
and the area of the surface $C_A^{(n)}$
\begin{align}
  \mathcal{A} &= \int_{C_A^{(n)}} d^{d-1}y \sqrt{g(y)}\,. \label{area}
\end{align}
Here $X^\mu(y)$ specify the embedding of the surface into the bulk, $y^i$ ($i=1,\dots,d-1$) the embedding coordinate,
and $g_{ij}(y)$ the induced metric on $C_n$,
\begin{align}
  g_{ij} (y) &= G_{\mu\nu}(X(y))
  \frac{\partial X^\mu}{\partial y^i}
  \frac{\partial X^\nu}{\partial y^j}\,.
\end{align}

The main difference from the Ryu-Takayanagi formula arises from the back-reaction of the codimension-two surface to the bulk metric $G_{\mu\nu}$.
Namely we extremize the action including the cosmic brane:
\begin{align}
  0\ =\frac{\delta I}{\delta G_{\mu\nu}(X)} =
  \frac{\delta I_\text{bulk}}{\delta G_{\mu\nu}(X)}
  + T_n\,\frac{\delta \mathcal{A}}{\delta G_{\mu\nu}(X)} \,,
\end{align}
where the first term is the original bulk equation of motion, and the second term is essentially the energy-momentum tensor of the cosmic brane $C_A^{(n)}$.
Note that $C_A^{(n)}$ is still a minimal surface as the equation of motion for the embedding $X^\mu(y)$ shows:
\begin{align}
  \frac{\delta I}{\delta X^\mu(y)}
  = T_n\,\frac{\delta \mathcal{A}}{\delta X^{\mu}(y)} = 0\,.
\end{align}
This equation should be evaluated on the backreacted bulk metric $G_{\mu\nu}$.
When there are matter fields $\psi$, we also have to solve
\begin{align}
  \frac{\delta I}{\delta \psi} =
  \frac{\delta I_\text{matters}}{\delta \psi} = 0\,,
\end{align}
in the backreacted background $G_{\mu\nu}$.
The Ryu-Takayanagi formula
\begin{align}
  S_A = \min_{\partial C_A = \partial A}
  \frac{\mathcal{A}[C_A]}{4G_N}\,,
\end{align}
is recovered from \eqref{HRE} in the limit $n\to1$ where the brane tension $T_n$ vanishes and we can neglect the backreaction of the brane.

\subsection{Derivation revisited from the viewpoint of the analogy}
To derive the holographic formula \eqref{HRE}, we employ the replica trick relating the trace of the density $\rho$ to the Euclidean partition function $Z$ \cite{Calabrese:2004eu},
\begin{align}
  \log\Tr[\rho^n] = \log Z[M_n] - n\log Z[M_1] \label{replica}\,,
\end{align}
where $M_n$ is the $n$-fold cover branched over the region $A$.
In the classical gravity regime, there exists a regular solution $B_n$ of the Einstein equation holographically dual to the field theory on the replica manifold $M_n$ such that $\partial B_n = M_n$.
The partition function $Z$ is equated to the on-shell bulk action on $B_n$:
\begin{align}
  Z[M_n] = Z_\text{bulk} \sim e^{-I_\text{bulk}[B_n]} \label{dual}\,.
\end{align}
$n$ has been supposed to be an integer up to now, but we analytically continue it to an arbitrary real number.
Such an analytic continuation can be performed in the bulk side by defining the ``bulk per replica'' manifold
\begin{align}
  \hat{B}_n = B_n / \mathbb{Z}_n,
\end{align}
under the assumption that the replica symmetry $\mathbb{Z}_n$ extends to the on-shell bulk solution $B_n$ \cite{Lewkowycz:2013nqa}.\footnote{
See \cite{Camps:2014voa} for the discussion on the replica symmetry $\mathbb{Z}_n$ breaking.
}
This quotient geometry $\hat{B}_n$ has a conical singularity at a codimension-two fixed locus $C_A^{(n)}$ of the $\mathbb{Z}_n$ symmetry with a deficit angle
\begin{align}
  \Delta\phi=2\pi(1-1/n)\,.
\end{align}
The fixed locus $C_A^{(n)}$ extends to the AdS boundary and touches on the entangling surface $\partial A$
which is also fixed locus of the replica symmetry.

Next, let us define ``bulk action per replica'' $I$ for the quotient $\hat{B}_n$,
just by dividing the bulk on-shell action $I_\text{bulk}[B_n]$ by $n$,
\begin{align}
  I \equiv I_\text{bulk}[B_n]/n\,. \label{I}
\end{align}
This action $I$ differs from $I_\text{bulk}[\hat{B}_n]$ of the quotient bulk $\hat{B}_n$,
and has an additional contribution from the singularity at $C_A^{(n)}$.\footnote{
Here our notation of $I_\text{bulk}[\hat{B}_n]$ is different from 
that in other literatures such as \cite{Dong:2016fnf}.
Our $I_\text{bulk}[\hat{B}_n]$ includes the contribution from the conical
singularity $C_A^{(n)}$, while their $I_\text{bulk}[\hat{B}_n]$ means $I_\text{bulk}[\hat{B}_n\backslash
C_A^{(n)}] = I_\text{bulk}[B_n]/n = I$
without the contribution from $C_A^{(n)}$.
}
Bearing in mind that $\hat{B}_n$ is locally the same as the original bulk $B_n$ away from the conical singularity $C_A^{(n)}$,
the Ricci scalar $R(X)$ of $\hat{B}_n$ takes the following form \cite{Fursaev:1995ef}
\begin{align}
  \sqrt{G(X)}R(X)|_{\hat{B}_n} =
  \sqrt{G(X)}R(X)|_{B_n}
  +2\Delta\phi\int_{C_A^{(n)}}d^{d-1}y\sqrt{g}\,\delta^{d+1}(X-X(y))
   \,. \label{ricci}
\end{align}
Thus in the Einstein gravity
$I_\text{EH} = -\frac{1}{16\pi G_N}\int d^{d+1}X\sqrt{G(X)}R(X) $,
\begin{align}
\begin{aligned}
  I_\text{bulk}[\hat{B}_n] &= I_\text{bulk}[B_n]/n - \frac{\Delta\phi}{8\pi G_N} \int_{C_A^{(n)}}d^{d-1}y \sqrt{g}\,,\\
  &=I-\frac{1-1/n}{4G_N}\mathcal{A}
\,, 
\end{aligned}
\end{align}
which means that the action $I$ includes the area term
\begin{align}
  I = I_\text{bulk}[\hat{B}_n]  + T_n\,\mathcal{A}\,, 
\end{align}
with the correct brane tension \eqref{tension}
\begin{align}
  T_n = \frac{\Delta\phi}{8\pi G_N}
   = \frac{1-1/n}{4G_N} \,,
\end{align}
and the area $\mathcal{A}$ \eqref{area} as desired.

A point of caution is that 
not $I_\text{bulk}[\hat{B}_n]$ itself, but the combination
$  I = I_\text{bulk}[\hat{B}_n] + T_n\,\mathcal{A}
$
is on-shell with respect to the bulk fields $G_{\mu\nu}(X)$ and $\psi(X)$.
This is clear from the relation \eqref{I} and $B_n$ being the regular solution for the action $I_\text{bulk}[B_n]$.

The replica symmetry would constrain the embedding $X^\mu(y)$ to be the minimal surface $\delta \mathcal{A }/\delta X^\mu(y)=0$.\footnote{
We could justify this statement somewhat by a following rough argument.
Consider how the area $\mathcal{A}$ would change in the leading order of a perturbation $\epsilon^\mu(y)$ of the embedding $X^\mu(y)$, in the bulk $\hat{B}_n$.
In the original bulk $B_n$,
where $n$ copies of $\hat{B}_n$ are glued at the surface,
let us call the vector $\epsilon^\mu(y)$
toward the $i$-th copy of $\hat{B}_n$
as $\epsilon^\mu_i(y)$.
Since the original surface $X^\mu(y)$ is invariant under the replica $\mathbb{Z}_n$ symmetry shifting
$\epsilon^\mu_i(y)$ to $\epsilon^\mu_{i+1}(y)$,
the variation of the area
$\frac{\delta \mathcal{A}}{\delta X^\mu} \epsilon_i^\mu$
does not depend on the label $i$ and in fact
$\frac{\delta \mathcal{A}}{\delta X^\mu} \epsilon_i^\mu
=\frac{\delta \mathcal{A}}{\delta X^\mu} \epsilon^\mu$.
On the other hand, the sum of these vectors vanishes
$\sum_{i=1}^n\epsilon_i^\mu=0$ because of the symmetry. 
Then $
  0 = \frac{\delta \mathcal{A}}{\delta X^\mu}\sum_{i=1}^n\epsilon_i^\mu
    = n \frac{\delta \mathcal{A}}{\delta X^\mu}\epsilon^\mu$ means that the area is minimal $\frac{\delta \mathcal{A}}{\delta X^\mu}=0$\,.
}
We promote the embedding $X^\mu(y)$ to a dynamical variable and minimize the action $I$ with respect to $X^\mu(y)$ in order to analytically continue $n$ to a real number.

Combining the replica trick \eqref{replica} and the holographic relation \eqref{dual} together with the definition of the action $I$ \eqref{I}, we have the expression
\begin{align}
\begin{aligned}
  \log\Tr[\rho^n] &= -(I_\text{bulk}[B_n] - n\, I_\text{bulk}[B_1]) \,,\\
  &= -n( I - I|_{n=1})\,,
\end{aligned}
\end{align}
from which the free energy $F(T)$ follows as the difference of the actions between $n$ and $n=1$
\begin{align}
  F &= -\frac{1}{n} \log \Tr[\rho^n]
  = I - I|_{n=1}\,.
\end{align}
The second term $-I|_{n=1}$ ensures the normalization $F(1)=-\log\Tr[\rho]=0$.
The free energy results from the minimization with respect to the fields
$\phi= \{ G_{\mu\nu}(X), \psi(X), X^\mu(y)\}$ 
\begin{align}
  F(T) &= \min_{\phi}\left(
    I[\phi]
  \right) - I|_{n=1}\,,
\end{align}
as the action $I$ is on-shell.
Here we introduce a temperature $T=1/n$ and rewrite the action as
\begin{align}
  I=  I_\text{bulk}[\hat{B}_n]+\left(1-T\right)\frac{\mathcal{A}}{4G_N}\,.
\end{align}
This succinct form is convenient to derive the entropy $\tilde{S}_n$
\begin{align}
  \tilde{S}_n &= -\frac{\partial F}{\partial T}
  =-\frac{\delta I[\phi]}{\delta\phi} \frac{\delta\phi}{\delta T} 
  +\frac{\mathcal{A}}{4G_N} \,,
\end{align}
where the first and second terms come from the variation of the fields $\phi$ and the tension $T_n=(1-T)/4G_N$, respectively.
Imposing the equations of motion, the first term vanishes $\delta I/\delta\phi=0$, and we reach the holographic R\'enyi entropy formula \eqref{HRE}
\begin{align}
  \tilde{S}_n 
  &=\frac{\mathcal{A}}{4G_N} \,.
\end{align}
The derivation explains why one has to take into account the backreaction of the cosmic brane to the geometry while extremizing the area.

Finally we derive the total energy $E$ by the Legendre transformation
\begin{align}
\begin{aligned}
  E &= F + T\tilde{S}_n \,,\\
    &= I_\text{bulk}[\hat{B}_n]-I_\text{bulk}[B_1]+\frac{\mathcal{A}}{4G_N}
     \,. \label{energy2}
\end{aligned}
\end{align}
This derivation is exactly the same as the one in thermodynamics;
$\delta E-T\delta S$ vanishes because of the minimization in the Legendre transformation $ F(T)\equiv\min_S(E(S)-TS)$, yielding
$\delta F=\delta(E-TS)=(\delta E-T\delta S)-S\delta T=-S\delta T$.
In our derivation of the holographic formula, the minimization of the free energy leads to a first-law like relation $0=\delta_\phi I=\delta_\phi E-T\delta_\phi \tilde{S}$.
The only difference is the meaning of the variation; $\delta_\phi$ is taken with respect to fields $\phi$ in our case.

\section{Proof of the R\'enyi entropic inequalities}\label{ss:Proof}
Having established the necessary tools in the preceding sections, we want to examine under what condition the holographic formula \eqref{HRE} satisfies the inequalities \eqref{inequality1}-\eqref{inequality4} of the R{\'e}nyi entropy.
Instead of dealing with the original ones, we prove the concise inequalities \eqref{inequality2_thermo}-\eqref{inequality4_thermo} whose physical meaning is more transparent.
They imply the stability of the system in the thermodynamic language, which is translated to the stability of the gravity theory as we will see soon.

\subsection{A holographic proof}
Some of the R\'enyi entropic inequalities follow straightforwardly from the holographic formula $\tilde{S}_n=\mathcal{A}/4G_N$ \eqref{HRE}.
The second inequality $\tilde{S}\ge0$ \eqref{inequality2_thermo} is trivial as the area $\mathcal{A}$ is always non-negative.
The non-negativity of the R\'enyi entropy $S_n=\frac{n}{n-1}F\ge0$,
which is equivalent to $F<0$ for $n<1$ and $F>0$ for $n>1$, also follows
from $\partial_n F=\tilde{S}_n/n^2\ge0$ and $F(1)=0$.
The first inequality \eqref{inequality1} descents from the forth inequality \eqref{inequality4_thermo} as mentioned in section \ref{ss:Stat}.

Let us move on to the proof of the forth inequality \eqref{inequality4_thermo}
\begin{align}
    C=-n\frac{\partial \tilde{S}_n}{\partial n} =
    -\frac{n}{4G_N}\frac{\delta \mathcal{A}}{\delta n} \ge0 \,. \label{C}
\end{align}
As the parameter $n$ varies slightly by $\delta n$,
the brane area $\mathcal{A}$ changes slightly by
\begin{align}
\begin{aligned}
  \frac{\delta \mathcal{A}[G,X]}{\delta n}
  &=\int d^{d+1}X\,
  \frac{\delta \mathcal{A}}{\delta G_{\mu\nu}(X)}
  \frac{\delta G_{\mu\nu}(X)}{\delta n}
  +\int d^{d-1}y\,
    \frac{\delta \mathcal{A}}{\delta X^{\mu}(y)}
  \frac{\delta X^{\mu}(y)}{\delta n}
\,,\\
  &=\int d^{d+1}X\,
  \frac{\delta \mathcal{A}}{\delta G_{\mu\nu}(X)}
  \frac{\delta G_{\mu\nu}(X)}{\delta n}\,,
\end{aligned}
\end{align}
where we used the minimality condition $\delta \mathcal{A}/\delta X^{\mu}=0$ for the embedding in the second equality.
Plugging this result into \eqref{C}, we have
\begin{align}
  C =
  -\frac{n}{4G_N} \int d^{d+1}X\,
  \frac{\delta \mathcal{A}}{\delta G_{\mu\nu}(X)}
  \frac{\delta G_{\mu\nu}(X)}{\delta n}\,. \label{C2}
\end{align}
The derivatives $\delta \mathcal{A}/\delta G_{\mu\nu}$ and $\delta G_{\mu\nu}(X)/\delta n$ are not independent due to the equation of motion of the bulk metric $G_{\mu\nu}$.
The variation with respect to $n$ gives
\begin{align}
  \frac{\delta I_\mathrm{bulk}}{\delta G_{\mu\nu}(X)}
    [G+\delta G,\psi+\delta\psi]
  + (T_n+\delta T_n)
      \frac{\delta \mathcal{A}}{\delta G_{\mu\nu}(X)}
        [G+\delta G,X+\delta X]
  = 0
\end{align}
or 
\begin{align}
  \frac{\delta I}{\delta G_{\mu\nu}(X)}
    [G+\delta G,X+\delta X,\psi+\delta\psi]
  + \frac{\delta n}{4 G_N n^2}
      \frac{\delta \mathcal{A}}{\delta G_{\mu\nu}(X)}
    [G,\psi,X]
  = 0\,,
\end{align}
where we used $\delta T_n=\delta n/(4G_Nn^{2})$.
In the leading order of $\delta n$,
the difference from the original equation motion is
\begin{multline}
  \int d^{d+1}X' \left[
  \frac{\delta^2 I}{\delta G_{\mu\nu}(X)\delta G_{\alpha\beta}(X')}
  \delta G_{\alpha\beta}(X')
  + 
  \frac{\delta^2 I}{\delta G_{\mu\nu}(X)\delta\psi(X')}
  \delta\psi(X')
  \right] \\
  + \int d^{d-1}y~
  \frac{\delta^2 I}{\delta G_{\mu\nu}(X)\delta X^\alpha(y)}
  \delta X^\alpha(y)
  + \frac{\delta n}{4G_Nn^2}
    \frac{\delta \mathcal{A}}{\delta G_{\mu\nu}}
  = 0\,.
\end{multline}
This gives the following relation
\begin{align}
  \frac{\delta \mathcal{A}}{\delta G_{\mu\nu}(X)}
  = -4G_Nn^2\int d^{d+1}X'
  \frac{\delta^2 I}{\delta G_{\mu\nu}(X)\delta G_{\alpha\beta}(X')}
  \frac{\delta G_{\alpha\beta}(X')}{\delta n}\,,
\end{align}
where we used the equations of motion $\delta I/\delta\psi=0$ and
$\delta I/\delta X^\mu=0$. 
Plugging this $\delta A/ \delta G$ into \eqref{C2}, finally we obtain a symmetric formula for the capacity of entanglement\footnote{
If we extend the domain of the integral from $\hat{B}_n$ to $B_n$ and use the action $I_\text{bulk}[B_n]=nI$, then the coefficient $n^3$ can be absorbed as 
\begin{align}\label{C_on_Bn}
  C
  &=\int_{B_n} d^{d+1}Xd^{d+1}X'~\frac{\delta G_{\mu\nu}(X)}{\delta n}
  \frac{\delta^2 I_\text{bulk}[B_n]}{\delta G_{\mu\nu}(X)\delta G_{\alpha\beta}(X')}
  \frac{\delta G_{\alpha\beta}(X')}{\delta n}\,.
\end{align}
This formula maybe applies to cases when the replica symmetry $\mathbb{Z}_n$ is spontaneously broken in the on-shell bulk $B_n$.
}
\begin{align}
  C
  &=n^3\int d^{d+1}Xd^{d+1}X'~\frac{\delta G_{\mu\nu}(X)}{\delta n}
  \frac{\delta^2 I}{\delta G_{\mu\nu}(X)\delta G_{\alpha\beta}(X')}
  \frac{\delta G_{\alpha\beta}(X')}{\delta n}\,. \label{heatcapacity}
\end{align}
To prove the non-negativity of $C$, it is sufficient to show that the Hessian
matrix $\frac{\delta^2 I}{\delta G_{\mu\nu}(X)\delta G_{\alpha\beta}(X')}$  is non-negative 
definite on the on-shell bulk $G_{\mu\nu}$. This condition means that the bulk geometry is stable against any perturbation, which is the main assumption in this paper as mentioned in Introduction. We will have a few comments on this assumption in section \ref{ss:Discussion}.

This proof also provides a holographic formula for calculating the capacity of entanglement $C$. Especially, the quantum fluctuation of the modular Hamiltonian
with respect to the original state is given by
\begin{align}
\begin{aligned}
  C(1) &= \braket{H^2} - \braket{H}^2 \,,\\
  &=\int d^{d+1}Xd^{d+1}X'~\frac{\delta G_{\mu\nu}(X)}{\delta n}
  \frac{\delta^2 I}{\delta G_{\mu\nu}(X)\delta G_{\alpha\beta}(X')}
  \frac{\delta G_{\alpha\beta}(X')}{\delta n}
  \Bigg|_{n=1} \,.
\end{aligned}
\end{align}

To prove the third inequality $E\ge0$ \eqref{inequality3_thermo}, we employ
the expression \eqref{energy2} and it is enough to show $I_\text{bulk}[\hat{B}_n]\ge
I_\text{bulk}[B_1]$ as $\hat{B}_n$ and $B_1$ obey the same boundary condition
$\partial \hat{B}_n=\partial B_1= M_1$. 
It is so since the functional $I_\text{bulk}$ is supposed to have a minimum
on the on-shell solution $B_1$, not the off-shell bulk $\hat B_n$, under the assumption that we can apply Gibbons-Hawking-Perry
prescription so that the Euclidean gravity action $I_\text{bulk}$ is non-negative definite.
Instead, we can derive the third inequality $E\ge0$
also from the second one $\tilde{S}_n\ge0$ and the fourth one $C\ge 0$,
in the same way as \cite{Hung:2011nu}. When $n\ge1$, the free energy $F$ is non-negative
because $\partial_n F=\tilde{S}_n/n^2\ge0$ and $F(1)=0$,
and so the energy $E=F+T\tilde{S}$ is also non-negative.
The non-negativity of the capacity $dE/dT=C\ge0$ means that the energy $E$ does not decrease with $T=1/n$
and is still non-negative even when $n\le1$.

\subsection{Legendre transformed expression for capacity of entanglement}
We derive another expression of the entanglement heat capacity \eqref{heatcapacity}
using the graviton propagator, following \cite{Chang:2013mca} which calculates holographic entanglement entropies with probe branes inserted in the bulk. 

We rewrite $\delta G_{\mu\nu}/\delta n$ appearing in \eqref{C2}, instead of $\delta \mathcal{A} /\delta  G_{\mu\nu}$.
By increasing the parameter $n$ slightly by $\delta n$,
the energy-momentum tensor of the brane
\begin{align}
  \bar{T}_{\mu\nu} &\equiv \frac{\delta I}{\delta G^{\mu\nu}} =\ \frac{\sqrt{G}}{2} T_{\mu\nu}
  = T_n\frac{\delta \mathcal{A}}{\delta G^{\mu\nu}} \,,
\end{align}
changes slightly as
\begin{align}
  \delta \bar{T}_{\mu\nu} &=
    \frac{1}{4G_N}
    \frac{\delta n}{n^2}
    \frac{\delta \mathcal{A}}{\delta G^{\mu\nu}} \,.
\end{align}
Correspondingly the bulk metric $G_{\mu\nu}$ shifts by
\begin{align}
\begin{aligned}
  \delta G_{\mu\nu}(X)
  &= 8\pi G_N\int d^{d+1}X'~
  G_{\mu\nu\alpha\beta}(X,X') ~
   2\delta \bar{T}^{\alpha\beta}(X') \label{propagator} \,,\\
  &= -4\pi\frac{\delta n}{n^2}\int d^{d+1}X'~
  G_{\mu\nu\alpha\beta}(X,X')
  \frac{\delta \mathcal{A}}{\delta G_{\alpha \beta}(X')} \,.
\end{aligned}
\end{align}
Here $G_{\mu\nu\alpha\beta}$ is the Green's function of the linearized Einstein equation on the fixed background $G_{\mu\nu}$.
Plugging it into \eqref{C2},
we obtain another expression of the entanglement heat capacity
\begin{align}
\begin{aligned}
  C &=
  \frac{\pi}{nG_N} \int d^{d+1}X d^{d+1}X'\,
  \frac{\delta \mathcal{A}}{\delta G_{\mu\nu}(X)}
  G_{\mu\nu\alpha\beta}(X,X')
  \frac{\delta A}{\delta G_{\alpha \beta}(X')} \,,\\
  &=
  \frac{1}{16nG_N^2} \int d^{d+1}X d^{d+1}X'\,
  \frac{\delta \mathcal{A}}{\delta G_{\mu\nu}(X)}
 \frac{\delta^2 \log Z[\bar{T}]}{\delta \bar{T}_{\mu\nu}(X) \bar{T}_{\alpha\beta}(X')}\bigg|_{\bar{T}=0}
  \frac{\delta A}{\delta G_{\alpha \beta}(X')}\,,
\end{aligned}
\end{align}
where $Z[\bar{T}]$
is the partition function with a source $\bar{T}_{\mu\nu}$ inserted.\footnote{
Here we assumed 
\begin{align}
  G_{\mu\nu\alpha\beta}(X,X') = 
  \frac{1}{16\pi G_N}\frac{\delta^2 \log Z[\bar{T}]}{\delta \bar{T}_{\mu\nu}(X)\delta \bar{T}_{\alpha\beta}(X')}\,,
\end{align}
which could be shown by taking the variation of 
$  \braket{G_{\mu\nu}(X)}_{\bar{T}} =
   {\delta \log Z[\bar{T}]}/{\delta \bar{T}_{\mu\nu}(X)} \,,
$ as
\begin{align}
  \delta \braket{G_{\mu\nu}(X)}_{\bar{T}} =
   \int d^{d+1}X'~\frac{\delta^{2} \log Z[\bar{T}]}{\delta \bar{T}_{\mu\nu}(X)\delta \bar{T}_{\alpha\beta}(X')} \delta\bar{T}_{\alpha\beta}(X')\,.
\end{align}
The normalization is determined  by the definition of the graviton propagator \eqref{propagator}\,.
}
In this form, the non-negativity of $C$ is equivalent to the concavity of $-\log Z[\bar{T}']$,
which holds for $-\log Z[\bar{T}]$
is a Legendre transformation of the bulk action $-\log Z[G_{\mu\nu}(X)] \simeq I[G_{\mu\nu}(X)]$ as
\begin{align}
  -\log Z[\bar{T}]
  = \min_{G_{\mu\nu}} \left(I[G_{\mu\nu}]-\int d^{d+1}X\, G_{\mu\nu}(X)\bar{T}^{\mu\nu}(X)\right)\,,
\end{align}
and in general the Legendre transformation $\mathcal{F}(J)\equiv\min_M[F(M)-JM]$ interchanges the convexity and the concavity, $\mathcal{F}''=-1/F''$.  

The explicit expression
\begin{align}
  \frac{\delta \mathcal{A}}{\delta G^{\mu\nu}(X)}
  &= -\frac{1}{2}\int d^{d-1}y \sqrt{g} \,
  g^{ij}
  \frac{\partial X^\mu}{\partial y^i}
  \frac{\partial X^\nu}{\partial y^j}
  \delta^{d+1}(X-X(y))\,,
\end{align}
allows us to rewrite the formula with integrals on the brane
\begin{align}
  C &=
  \frac{\pi}{4G_N n} \int d^{d-1}y\, d^{d-1}y'\,
  \sqrt{g(y)}\sqrt{g(y')}
  \frac{\partial X^\mu}{\partial y^i}
  \frac{\partial X^\nu}{\partial y_i}
  G_{\mu\nu\alpha\beta}(X(y),X(y'))
  \frac{\partial X^\alpha}{\partial y'^j}
  \frac{\partial X^\beta}{\partial y'_j}\label{HolC}\,.
\end{align}

This representation is a consequence of the Legendre transformation between the response $G_{\mu\nu}$ and the source $\bar{T}^{\mu\nu}$.
In fact,
for a free energy $F(M_i)$
with general responses $M_i$ such as magnetization or chemical potential, the dual free energy $\mathcal{F}(J^i)$
\begin{align}
  \mathcal{F}(J^{i}) = \min_{M_i} [F(M_i)-J^iM_i]\,,
\end{align}
with $J^i$ the dual sources such as magnetic field or charge,
satisfies
\begin{align}
  \delta M_i
  \frac{\partial^2 F}{\partial M_i\partial M_j}
  \delta M_j
  &=\delta J^i \delta M_i
  =-\delta J^i
  \frac{\partial^2 \mathcal{F}}{\partial J^i \partial J^j}
  \delta J^j\,,
\end{align}
as $\delta F=J^i\delta M_i$ and 
$\delta \mathcal{F}=-M_i\delta J^i$.
The Legendre transformation
interchanges the convexity and the concavity.

\section{Calculations of the capacity of entanglement}\label{ss:Capacity}
Our holographic proof of the inequalities for the R{\'e}nyi entropy highlights a role of the stability in the bulk as a unitarity of the dual field theory.
The discussion was illuminating for the formal proof, but less concrete so far.
In this section, we switch gears and move onto tangible examples of the capacity of entanglement in various systems.

\subsection{Conformal field theory}

In two-dimensional CFT with central
charge $c$ the R{\'e}nyi entropies for an interval of length $L$ are well-known \cite{Holzhey:1994we,Calabrese:2004eu}
\begin{align}
        S_n = \frac{c}{6}\left( 1 + \frac{1}{n} \right) \log (L/\epsilon)
\,,
\end{align}
with the UV cutoff $\epsilon$.
It yields the capacity of entanglement straightforwardly
\begin{align}\label{2d_CFT_Cn}
        C(n) = \frac{c}{3n}\log (L/\epsilon) \,.
\end{align}
As it shows, the capacity is always positive in accord with the inequality \eqref{inequality4_thermo}
as the length $L$ cannot be smaller than the UV cutoff $\epsilon$.

It is challenging to obtain the capacity $C(n)$ for general $n$ in higher dimensional CFT, while one can calculate $C(n)$ of a sphere in the limit $n\to1$.
This is because $C(1)=-\partial_n\tilde{S}_n|_{n=1}$ is identical to the derivative of the R\'enyi entropy $C(1) =
-2\partial_n S_n|_{n=1}$,
whose calculations were already carried out for a sphere in CFT in \cite{Perlmutter:2013gua}.
In this case, the capacity becomes
\begin{align}
  C(1) &= \text{Vol} (\mathbb{H}^{d-1})\,
    \frac{2\pi^{d/2+1}(d-1)\Gamma(d/2)}{\Gamma(d+2)}C_T\,.\label{C(1)}
\end{align}
This is proportional to the coefficient $C_T$ of the correlation function of
the energy-momentum tensor \cite{Osborn:1993cr}
\begin{align}
  \braket{T_{ab}(x)T_{cd}(0)} = C_T \frac{I_{ab,cd}(x)}{x^{2d}},\, 
\end{align}
where $I_{ab,cd}(x)$ is a function given by
\begin{align}
\begin{aligned}
  I_{ab,cd}(x) &=
    \frac{1}{2}\left(I_{ac}(x)I_{bd}(x) + I_{ad}(x)I_{bc}(x)\right)
    - \frac{1}{d}\delta_{ab}\delta_{cd} \,,\\
  I_{ab}(x)\ &= \delta_{ab} - 2\frac{x_ax_b}{x^2}\,.
\end{aligned}
\end{align}

The positivity of $C(1)$ manifests itself in the form \eqref{C(1)} as the
volume of the hyperbolic space is positively divergent.
In practice, it is convenient to introduce the regularized volume\footnote{To
derive \eqref{Reg_Volume}, one can either put a cutoff near the infinity
of the hyperbolic space, or use a dimensional regularization.
In the former case, one ignores the power-law divergences for the cutoff
to extract the universal part, while in the latter case one analytically
continues the dimension $d$ from the range $1< d < 2$ to an arbitrary value.
}
\begin{align}\label{Reg_Volume}
        \text{Vol} (\mathbb{H}^{d-1}) = \pi^{d/2 -1} \Gamma\left( 1 - \frac{d}{2}\right)
\,,
\end{align}
to read off the so-called universal part of the R{\'e}nyi entropies.
This operation corresponds to adding local counter terms with respect to
the background metric to render the partition function finite.
It works well for any $d$ except even integers as the poles structure of the gamma function shows in \eqref{Reg_Volume}.
This signals the Weyl anomaly that cannot be removed by local counter terms.
In even $d$ dimensions, one has to replace the formula \eqref{Reg_Volume}
with \cite{Casini:2011kv,Hung:2011nu}
\begin{align}\label{Reg_Volume_even}
        \text{Vol} (\mathbb{H}^{d-1}) = \frac{2 (-\pi)^{d/2 -1}}{\Gamma(d/2)}\log
(R/\epsilon)  \,, \qquad (d: \text{even}) \,,
\end{align}
by introducing the UV cutoff $\epsilon$ and the radius $R$ of the hyperbolic
space.
When applied to the entropy of an interval of width $L$ in $d=2$, the radius
of the hyperbolic space $R$ should be identified with the width $L/2$ in the
regularized volume \eqref{Reg_Volume_even} and we are able to recover the CFT$_2$
result \eqref{2d_CFT_Cn} upon the relation $C_T = c/(2\pi^2)$.

\subsection{Free fields}
The capacity of entanglement is less tractable to calculate for interacting QFTs
as the modular Hamiltonian is non-local in general.
For free field theories, things are much simpler and one is able to compute the R{\'e}nyi entropies
using the partition function on $\BS^1\times \BH^{d-1}$ which is conformally
equivalent to the replica space of a spherical entangling surface \cite{Casini:2010kt,Casini:2011kv,Klebanov:2011uf}
(see also \cite{Dowker:2010bu,Dowker:2010yj,Solodukhin:2010pk}).

Firstly we consider a conformally coupled real massless scalar field.
With the help of the map to $\BS^1\times \BH^{d-1}$, the partition function on the $n$-fold replica manifold of a spherical entangling surface becomes \cite{Klebanov:2011uf}
\begin{align}
        \log Z_s(n) = - \int_0^\infty d\lambda\, \mu_s (\lambda)\, \left[
\log\left( 1- e^{-2\pi n \sqrt{\lambda}}\right) + \pi n \sqrt{\lambda} \right]
\,,
\end{align}
where $\mu_s(\lambda)$ is the Plancherel measure of the scalar field on $\BH^{d-1}$
\cite{Camporesi:1990wm,Bytsenko:1994bc}
\begin{align}
        \mu_s (\lambda) = \frac{\text{Vol}(\BH^{d-1})}{2^{d-1} \pi^{\frac{d+1}{2}}
\Gamma\left(\frac{d-1}{2}\right)} \sinh (\pi \sqrt{\lambda})\left| \Gamma\left(
\frac{d}{2} -1 + i \sqrt{\lambda}\right)\right|^2 \,.
\end{align}
Together with \eqref{positivity_of_c}, it leads to the capacity of entanglement 
\begin{align}\label{Cs}
        C_s (n) = \pi^2 n^2 \int_0^\infty d\lambda\, \mu_s(\lambda)\, \lambda
\,\text{csch}^{2} \left(\pi n \sqrt{\lambda}\right) \,.
\end{align}

Turning into a massless Dirac fermion, the partition function is written
as
\begin{align}
        \log Z_f(n) = \int_0^\infty d\lambda\, \mu_f (\lambda)\, \left[ \log\left(
1+ e^{-2\pi n \lambda}\right) + \pi n \lambda \right] \,,
\end{align}
where the Plancherel measure of the spinor on $\BH^{d-1}$ is \cite{Bytsenko:1994bc}
\begin{align}
        \mu_f(\lambda) = \frac{g(d)\, \text{Vol}(\BH^{d-1})}{2^{d-2} \pi^{\frac{d+1}{2}}
\Gamma\left(\frac{d-1}{2}\right)} \cosh(\pi\lambda)\left| \Gamma\left( \frac{d-1}{2} + i \lambda\right)\right|^2\
,
\end{align}
and $g(d)\equiv 2^{[ d/2]}$ is the dimension of Dirac spinors in $d$ dimensions.
The capacity takes a similar form to the scalar field:
\begin{align}\label{Cf}
        C_f (n) = \pi^2 n^2 \int_0^\infty d\lambda\, \mu_f(\lambda)\, \lambda^2 
\,\text{sech}^{2} \left(\pi n \lambda\right) \,.
\end{align}
Both \eqref{Cs} and \eqref{Cf} are manifestly positive in their forms.

In two dimensions $d=2$, these capacities reproduces the CFT$_2$ result \eqref{2d_CFT_Cn} with $c=1$. 
They are also consistent with the general formula \eqref{C(1)} of
$C(1)$ for CFT where the free fields have the following values of $C_T$ \cite{Osborn:1993cr}
\begin{align}
        (C_T)_\text{scalar} = \frac{d\, \Gamma(d/2)^2}{4\pi^d (d-1)} \,,
\qquad (C_T)_\text{fermion} = \frac{g(d)\, d\, \Gamma(d/2)^2}{8\pi^d} \,.
\end{align}

For massive cases and for a region $A$ other than a ball,
it is hard to obtain capacities analytically,
 but we can resort to lattice discretization to calculate them numerically.
The partition functions $\Tr[\rho_A^n]$ are expressed  by correlation functions of discretized fields located in the region $A$ as follows \cite{Srednicki:1993im,Peschel03,Casini:2009sr}.
For free scalars $\phi_i$ and its conjugates $\pi_i$ with correlation functions $X_{ij}=\braket{\phi_i\phi_j}$ and $P_{ij}=\braket{\pi_i\pi_j}$,
the partition function is given by
\begin{align}
  \log\Tr[\rho^n_A] &= -\Tr\left[\log\Big((D_{s}+1/2)^n-(D_{s}-1/2)^n\Big)\right]
  = -\sum_a\Big(\log(e^{n\epsilon_a}-1) - n\log(e^{\epsilon_a}-1)\Big)\,,
\end{align}
where we set the eigenvalues of $D_s=\sqrt{XP}\,(\ge1/2)$ as $\coth(\epsilon_a/2)/2$.
The indices $i,j$ run only the ones corresponding to the sites inside the region $A$.
This yields a manifestly non-negative capacity
\begin{align}
  C_s(n) = n^2\Tr\left[\frac{(D_s+1/2)^n(D_s-1/2)^n}{((D_s+1/2)^n-(D_s-1/2)^n)^2}
  \left(\log\frac{D_s+1/2}{D_s-1/2}\right)^2\right]
  =\frac{n^2}{4}\sum_a \epsilon_a^2\,\text{csch}^2(n\epsilon_a/2)\,.
\end{align} 
The calculation for free fermions $\psi_i$ is similar \cite{PhysRevLett.105.080501}.
The partition function given by
\begin{align}
  \log\Tr[\rho^n_A] &= \Tr\left[\log\Big((1-D_f)^n+D_f^n\Big)\right]
  = \sum_a\Big(\log(e^{n\epsilon_a}+1) - n\log(e^{\epsilon_a}+1)\Big)\,,
\end{align}
 yields a manifestly non-negative capacity
\begin{align}
  C_f(n) = n^2\Tr\left[\frac{D_f^n(1-D_f)^n}{(D_f^n+(1-D_f)^n)^2}
  \left(\log\frac{D_f}{1-D_f}\right)^2\right]
  =\frac{n^2}{4}\sum_a \epsilon_a^2\,\text{sech}^2(n\epsilon_a/2)\,,
\end{align} 
where the eigenvalues of the matrix $(D_f)_{ij}=\braket{\psi_i\psi_j^\dag}$ are $1/(1-e^{\epsilon_a})$.

\subsection{Gravity duals}
The R{\'e}nyi entropies of a spherical entangling surface are calculated
through the holography using the AdS topological black hole \cite{Hung:2011nu}.
The metric for the bulk per replica $\hat{B}_n$ is known to be 
\begin{align}
  ds_{d+1}^2 = \frac{dr^2}{f_{n}(r)} + f_n(r)d\tau^2 +r^2(du^2+\sinh^2u\, d\Omega^2_{d-2})\label{metric}\,,
\end{align}
with a function
\begin{align}
  f_{n}(r) = r^2 - 1 -\frac{r_n^d - r_n^{d-2}}{r^{d-2}}\,.
\end{align}
The Euclidean time direction $\tau$ has the period $\tau \sim \tau + 2\pi$
so that this metric reduces to the non-singular flat space
$d\tau^2+du^2+\sinh^2u\,d\Omega^2_{d-2} \sim \sum_{i=1}^{d} dx_i^2$
at the conformal boundary $r\to\infty$.
This geometry has a conical singularity $C_A^{(n)}$ at the horizon $r=r_n$ ($f_n(r_n)=0$), where the Euclidean time $\tau$ circle shrinks to a point. The horizon radius $r_n$ is determined by $n$ as
\begin{align}
  \frac{2\pi}{n} = \frac{f'_n(r_h)}{2}2\pi
  \quad&\Leftrightarrow \quad
   n = \frac{2}{f'_n(r_h)}
  = \frac{2}{dr_n-(d-2)r_n^{-1}} \label{r_n}\\
   &\Leftrightarrow\quad
   r_n = \frac{1 + \sqrt{1 + n^2 d (d-2)}}{n\, d}\,,
\end{align}
such that the correct
conical singularity $\tau\sim\tau+2\pi/n$ is reproduced.
$r_n$ is monotonically decreasing with $n$ and satisfies $r_n\ge \lim_{n\to\infty}r_n=\sqrt{(d-2)/d}$.

The cosmic brane is located on the horizon and 
the improved R{\'e}nyi entropy is nothing but the black hole entropy
\begin{align}
  \tilde{S}_n 
  =r_n^{d-1}\frac{\text{Vol}(\mathbb{H}^{d-1})}{4G_N}\,.
\end{align}
Integration by $n$ gives the free energy $F(n)$
\begin{align}
  F(n) &=\int_1^ndn'\,\frac{\tilde{S}_{n'}}{n'^2}
  = \frac{\text{Vol}(\mathbb{H}^{d-1})}{4G_N}
    \frac{2-r_n^d-r_n^{d-2}}{2}\,,
\end{align}
where we used the relation
$(d+(d-2)/r_n^2)\partial_nr_n=-2/n^2$ followed from the expression \eqref{r_n}.
This means that the R\'enyi entropy $S_n=n F/(n-1)$ is
\begin{align}
  S_n 
  &= \frac{n}{n-1}\frac{\text{Vol}(\mathbb{H}^{d-1})}{4G_N}
    \frac{2-r_n^d-r_n^{d-2}}{2}\,,
\end{align}
which is non-negative for any $n$ and $d$ as $r_n>1$ for $n<1$
and $r_n<1$ for $n>1$.
We can also check that the first inequality \eqref{inequality1} holds or equivalently
$S[\rho_n|\rho]=-(n-1)^2\partial_n S_n=1+ (d-1)(r_n^d-r_n^{d-2})/2 -r_n^{d-1}\ge0$.
The total energy \eqref{energy2} and the capacity of entanglement \eqref{C} given by
\begin{align}
  E &= \frac{\text{Vol}(\mathbb{H}^{d-1})}{4G_N}
    \frac{2+(d-1)(r_n^d-r_n^{d-2})}{2} \,,\\
 C &= 
  \frac{\text{Vol}(\mathbb{H}^{d-1})}{4G_N} (d-1)\,r_n^{d-1}\frac{d\,r_n^2-(d-2)}{d\,r_n^2+(d-2)}\label{HolC2}
  \,,
\end{align}
are also non-negative for any $n$ and $d$ as $r_n>1$ for $n<1$
and $r_n\ge\sqrt{(d-2)/d}$ for $n>1$.

A more direct way to get $C$ without knowing $\tilde S_n$ is to use the formula
\begin{align}\label{Cformula2}
  C =
  -\frac{n}{8G_N} \int_{C_A^{(n)}} d^{d-1}y \sqrt{g}\,
  g^{ij}\,
  \frac{\partial X^\mu}{\partial y^i}
  \frac{\partial X^\nu}{\partial y^j}\,
  \frac{\delta G_{\mu\nu}(X(y))}{\delta n}\,,
\end{align}
which is equivalent to the previous ones \eqref{C2} and \eqref{HolC}.\footnote{
Even when the graviton propagator $G_{\mu\nu\alpha\beta}(X,X')$ is known, the expression \eqref{HolC} is too difficult to evaluate in general and it suffers from a subtle contribution from the asymptotic boundary.
We will comment on this difficulty in Appendix \ref{ap:Graviton}.
}
When applying \eqref{Cformula2} to the background \eqref{metric}, we take the embedding $X^\mu(y)$ of the surface as
$(X^r, X^\tau, X^i) = (1, 0, y^i)$,
where $y^i$ are the coordinates of $\mathbb{H}^{d-1}$.
For $\mathbb{H}^{d-1}$ is maximally symmetric, the integration just gives its volume and the formula reads
\begin{align}
  C &=
  -\frac{n\text{Vol}(\mathbb{H}^{d-1})}{8G_N} r_n^{d-3}\,
  \frac{\delta G_{uu}}{\delta n}\bigg|_{C_A^{(n)}}\,.
\end{align}
Reassuringly it agrees with \eqref{HolC2} as $\delta G_{uu}/\delta n = \delta r_n^2/\delta n
  = 2 r_n \partial_n r_n$.

When $n=1$, the holographic capacity of entanglement takes a particularly simple form
\begin{align}
  C(1) =
  \frac{\text{Vol}(\mathbb{H}^{d-1})}{4G_N}\,. \label{holC(1)}
\end{align}
It takes exactly the same form as the field theory calculation \eqref{C(1)} because the holographic system has \cite{Buchel:2009sk}
\begin{align}
  C_T &= \frac{1}{8\pi G_N}\frac{d+1}{d-1}\frac{\Gamma(d+1)}{\pi^{d/2}\Gamma(d/2)}\,.
\end{align}

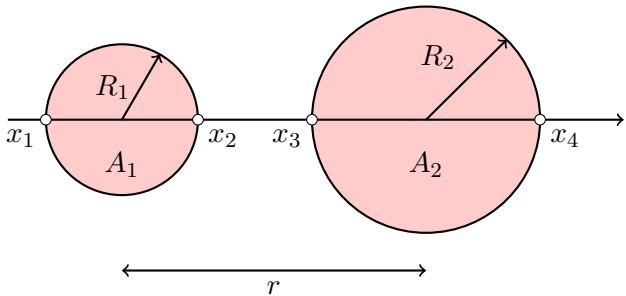
\begin{figure}[htbp]
\centering
\begin{tikzpicture}
        \begin{scope}
        \draw [name path=circleone, thick, fill=red!20] (-1,0) circle (1cm);
        \draw [name path=circletwo, thick, fill=red!20] (3,0) circle (1.5cm);
        \draw [name path=line, thick, ->, fill=white] (-2.5,0) -- (5.6,0)
node[right] {};
        \draw [thick, ->] (-1,0) -- ++(60:1cm) node[midway, left] {$R_1$};
        \draw [thick, ->] (3,0) -- ++(45:1.5cm) node[midway, above left]
{$R_2$};
        \draw [name intersections={of=circleone and line, by={x2, x1}}, fill=white]

        (x2) circle (2pt) node [below right] at (x2) {$x_2$}
        (x1) circle (2pt) node [below left] at (x1) {$x_1$};
        \draw [name intersections={of=circletwo and line, by={x4, x3}}, fill=white]
        (x3) circle (2pt) node [below left] at (x3) {$x_3$}
        (x4) circle (2pt) node [below right] at (x4) {$x_4$};
                \node  at (-1,-0.3) [below] {$A_1$};
        \node  at (3,-0.3) [below] {$A_2$};
        \draw[thick, <->] (-1, -2) -- (3, -2) node[midway, below] {$r$};
        \end{scope}
\end{tikzpicture}
\caption{The entangling region (shown in red) consists of two balls $A_1$ and $A_2$ of radii $R_1$ and $R_2$, respectively.
The four coordinates $x_i$ are defined on the line connecting the
centers of the balls. In a conformal field theory, the configuration of the balls
is uniquely specified by the cross ratio $x\equiv(x_1-x_2)(x_3-x_4)/(x_1-x_3)(x_2-x_4)=4R_1R_2/(r^2-(R_1-R_2)^2)$,
where $r$ is the distance between the two centers.}
\label{fig:TwoBalls}
\end{figure}

One more example we are going to show is the system with two balls $A_1$ and $A_2$  of radii $R_1$ and $R_2$ separated enough (see Fig.\,\ref{fig:TwoBalls}).
The R{\'e}nyi entropy of the two balls for an arbitrary $n$ is beyond our scope, but a perturbative calculation is feasible in the leading linear order of $\delta n\equiv n-1$.
Indeed, an analog of the mutual information $I^{(n)}(A_1,A_2)\equiv S_n(A_1)+S_n(A_2)-S_n(A_1 \cup A_2)$ has been evaluated holographically by \cite{Dong:2016fnf} for $n$ close to $1$.
We will benefit from the result to get the capacity of entanglement $C_{A_1\cup A_2}$ for the union of the two balls $A_1$ and $A_2$ in this parameter region.

The positions of the balls are parametrized by the cross ratio
$0\le x\equiv \frac{(x_1-x_2)(x_3-x_4)}{(x_1-x_3)(x_2-x_4)}\le1$\,.
$x_i$ are the coordinates of the points where the line connecting the two centers intersects the balls, $x_{1,2}$ for $A_1$ and $x_{3,4}$ for $A_2$ (Fig.\,\ref{fig:TwoBalls}).
There are two phases depending on the topology of the minimal surfaces in the bulk, and there is a critical point $x =x_c$ below which a disconnected surface is favored, otherwise a connected one is realized \cite{Headrick:2010zt}.
The calculation of $I^{(n)}(A_1,A_2)$ performed by \cite{Dong:2016fnf} is in the disconnected phase ($x\le x_c$) with the balls separated enough.
To convert the result into the capacity $C(1)$, 
we apply a derivative $-2\partial_n|_{n=1}$ on $S_n(A_1 \cup A_2)=S_n(A_1)+S_n(A_2)-I^{(n)}(A_1,A_2)$ to get
\begin{align}
  &C_{A_1 \cup A_2}(1) = C_{A_1}(1) + C_{A_2}(1) + 2\partial_nI^{(n)}(A_1,A_2)|_{n=1} \\
  &= \frac{\text{Vol}(\mathbb{H}^{d-1})_{R_1}+\text{Vol}(\mathbb{H}^{d-1})_{R_2}}{4G_N}   + \frac{2^{4-d}\pi^{d+1}C_T}{d(d^2-1)\Gamma\left( (d-1)/2 \right)^2}
  \frac{2-x}{x}B
  \left(\left(\frac{x}{2-x}\right)^2;\frac{d+1}{2};\frac{2-d}{2}\right)\,,
\end{align}
where $\text{Vol}(\mathbb{H}^{d-1})_{R}$ is the regularized volume of $\mathbb{H}^{d-1}$ of radius $R$ given by \eqref{Reg_Volume} and \eqref{Reg_Volume_even}.

\subsection{Large and small $n$ limits}
Before closing this section, we examine the large and small $n$ behaviours of the capacity $C(n)$
for a spherical entangling region in the systems we have studied.
In the thermodynamic interpretation, we regard these as the low and high temperature limits for the temperature $T=1/n$.

In the low temperature limit $n\to \infty$,
 the capacities of conformal theories go to zero as
\begin{align}
        C_s (n) ~
&\sim ~
\text{Vol}(\BH^{d-1})\,\frac{\Gamma(
\frac{d}{2} -1)^2}{15\cdot 2^{d-1} \pi^{\frac{d-3}{2}}
\Gamma\left(\frac{d-1}{2}\right)}\frac{1}{n^3} \quad(d\neq2)\,, \\
        C_f (n) ~&\sim~ \text{Vol}(\BH^{d-1})\,\frac{g(d))\Gamma(\frac{d-1}{2})}{3\cdot 2^d \pi^\frac{d-1}{2}}\frac{1}{n}\,, \\
  C_\text{AdS}(n) ~&\sim~
  \frac{\text{Vol}(\mathbb{H}^{d-1})}{4G_N}
  \frac{(d-1)(d-2)^{d/2-1}}{d^{d/2}}\frac{1}{n}\,,
\end{align}
for the massless scalar, massless fermion and CFT dual to the AdS spacetime, respectively.
They are proportional to a power of the temperature $T=1/n$, indicating
a gapless excitation for the modular Hamiltonian.
In $d=2$, the scalar capacity also becomes proportional to $1/n$ as $C_s(n) = (c/3n) \log(L/\epsilon)$.

On the other hand,
in the high temperature limit $n\to0$,
they obey the Stefan-Boltzmann's law $C(T)\propto T^{d-1}$
for thermal massless gases
\begin{align}
        C_s (n) ~&\sim~ \text{Vol}(\BH^{d-1})\,\frac{(d-1)\Gamma(d/2+1)\zeta(d)}{2^{d-2}
\pi^{\frac{3}{2}d-1}
}\frac{1}{n^{d-1}}\,, \\ 
        C_f (n) ~&\sim~
        \text{Vol}(\BH^{d-1})\,\frac{(d-1)(2^{d-1}-1)\Gamma(d/2+1)\zeta(d)g(d)\,}{2^{2d-3}
\pi^{\frac{3}{2}d-1}}\frac{1}{n^{d-1}}
       \,, \\ 
  C_\text{AdS}(n) ~&\sim~
  \frac{\text{Vol}(\mathbb{H}^{d-1})}{4G_N}
  (d-1)\left(\frac{2}{nd}\right)^{d-1}\,.
\end{align}
To derive these results,
we used asymptotic behavior of $\mu(\lambda)$
\begin{align}
  \mu_s(\lambda) &\sim \frac{\text{Vol}(\BH^{d-1})}{2^{d-1} \pi^{\frac{d-1}{2}}
\Gamma\left(\frac{d-1}{2}\right)} \lambda^{\frac{d-3}{2}} \,, &
        \mu_f(\lambda) &\sim \frac{g(d)\, \text{Vol}(\BH^{d-1})}{2^{d-2} \pi^{\frac{d-1}{2}}
\Gamma\left(\frac{d-1}{2}\right)} \lambda^{d-2}
,
\end{align}
in the limit $\lambda \to \infty$ and mathematical relations
\begin{align}
  \int_0^\infty dx~x^d\text{csch}^2x &= \frac{\Gamma(d+1)\zeta(d)}{2^{d-1}} \,, \\
  \int_0^\infty dx~x^d\text{sech}^2x &= \frac{(2^{d-1}-1)\Gamma(d+1)\zeta(d)}{4^{d-1}} \,,
\end{align}
and $\Gamma(d+1)/2^d=\Gamma(\frac{d+1}{2})\Gamma(\frac{d}{2}+1)/\sqrt{\pi}$.

\section{Discussion}\label{ss:Discussion}
Our approach to the holographic R{\'e}nyi entropy is advantageous for formal proofs and provides a clear-cut relation of the roles played by the unitarity in QFT and the stability of the gravity theory.
Meanwhile, the holographic formula lacks a power of computability in a practical problems as we saw in section \ref{ss:Capacity}.
The main difficulty originates from the procedure of finding the extremal surface of a cosmic brane in the backreacted geometry.
One would be able to calculate the R{\'e}nyi entropy perturbatively either in $n-1$ or in shape, otherwise it is generically unattainable in its nature.
It is still algorithmically simple to implement in numerical calculation that would be worth more investigation.

We do not know any rigorous proof or plausible argument for the bulk stability against any perturbation that is essential in our holographic proof of the inequalities.
To answer a question whether the bulk is stable or not requires the knowledge of quantum gravity which remains to be developed.
It is one of the fundamental problems even in the perturbative Euclidean quantum gravity and providing the complete solution is far beyond the scope of this paper.
We comment on possible attempts instead:
\begin{itemize}
\item
The assumption we made for the bulk stability is a sufficient condition, but may not be a necessary condition, to prove the R\'enyi entropic inequalities in the holographic system.
Namely the non-negativity of the heat capacity \eqref{heatcapacity} could have followed from the condition for the Hessian matrix to be non-negative definite only in the subspace of the metric variation $\delta G_{\mu\nu}/\delta n$ induced by changing the replica parameter.
Unfortunately we were not able to demonstrate the non-negativity of the Hessian in the subspace as the metric variation is only calculable in the neighbourhood of the cosmic brane.
\item 
The perturbative Euclidean gravity is known to suffer from the bulk instability due to the Weyl mode.\footnote{We
thank M.\,Headrick for drawing our attention to this subtlety.
}
There are at least two directions known in literature to fix this problem:
one (ad-hoc) attempt is Gibbons-Hawking-Perry prescription which claims to change the contour of integration for the Weyl mode, called conformal rotation, in the path integral formulation of the perturbative Euclidean gravity \cite{Gibbons:1978ac,Christensen:1979iy,Hawking:1980gf}. (See also \cite{Mazur:1989by,Mottola:1995sj} for further discussions.)
For locally Euclidean AdS$_3$ spaces, this prescription gives the correct one-loop partition function of gravity expected from the AdS$_3$/CFT$_2$ correspondence \cite{Giombi:2008vd}, 
and it might well work for more general holographic theories at the one-loop level.
The other is based on the canonical quantization of gravity to show the Hamiltonian is bounded from below, and then continues to Euclidean path integral with an appropriate choice of contour \cite{Arisue:1986pw,Schleich:1987fm}.
The two approaches appear to be complimentary to each other, but a precise relation between them has not been completely explored.
\end{itemize}

As a future direction, it would also be intriguing to include quantum corrections to the holographic R{\'e}nyi entropy \cite{Faulkner:2013ana}.
Recent discussions \cite{Jafferis:2014lza,Jafferis:2015del} argues a relation between the boundary modular Hamiltonian $H_\text{bdy}$ and the bulk one $H_\text{bulk}$
\begin{align}
        H_\text{bdy} = \frac{\hat A}{4G_N} + H_\text{bulk} + \hat S_\text{Wald-like} + O(G_N) \,.
\end{align}
Here $\hat A$ is an operator in the bulk which is supposed to give the area of the Ryu-Takayanagi surface $\CS$ when sandwiched by a state dual to a given state in the boundary field theory.
$\hat S_\text{Wald-like}$ denotes local operators localized on $\CS$ in the semi-classical limit.
It may as well be applied to the calculation of the capacity \eqref{positivity_of_c} for $n=1$, leading to
\begin{align}
        C(1)_\text{bdy} = \frac{1}{16G_N^2}\left( \langle \hat A^2 \rangle - \langle \hat A \rangle^2\right) + \frac{1}{2G_N}\left( \langle \hat A\, \tilde H_\text{bulk} \rangle - \langle \hat A \rangle \langle \tilde H_\text{bulk} \rangle \right) + O(1) \,,
\end{align}
where we introduced $\tilde H_\text{bulk}\equiv H_\text{bulk} + \hat S_\text{Wald-like}$ to simplify the notation.
Surprisingly, the leading term is of order $1/G_N^{2}$, which was not observed in the examples in section \ref{ss:Capacity}.
Thus we are lead to conclude that the area operator has to satisfy
\begin{align}
        \alpha\equiv \frac{\langle \hat A^2 \rangle - \langle \hat A \rangle^2}{8 G_N} = O(G_N^0) \,.
\end{align}
We believe this is a defining property of the area operator that holds for any state in the semi-classical limit.
A similar statement has been made in, e.g., \cite{Almheiri:2016blp}
in the context of the linearity of the area operator recently.
The order $1/G_N$ term is likely to contribute to the capacity, and it indeed does so for the cases considered in section \ref{ss:Capacity}.
We do not know how to estimate it in practice, but the non-negativity of the capacity yields a constraint 
\begin{align}
        \alpha + \langle \hat A\, \tilde H_\text{bulk} \rangle - \langle \hat A \rangle \langle \tilde H_\text{bulk} \rangle \,\ge\, 0 + O(G_N)\,.
\end{align}
Testing this inequality needs more detailed information on the area operator and the local operators on the Ryu-Takayanagi surface $\CS$, which is far beyond the scope of the present work.

Another interesting direction is to generalize the holographic formula of the R{\'e}nyi entropy to a time dependent background \cite{Hubeny:2007xt} and higher derivative gravities \cite{Dong:2013qoa,Camps:2013zua}.
It is not so obvious how a cosmic brane modifies the original proposals, but it is likely that the entropy is still given by variants of the area formula.


\acknowledgments  
We are grateful to T.\,Kawano, Y.\,Nakagawa and Y.\,Sato for valuable discussions. 
We also thank the Yukawa Institute for Theoretical Physics for hospitality during the completion of the paper.
The work of YN was supported by World Premier International Research Center Initiative (WPI), MEXT, Japan.
The work of YN was also supported in part by JSPS Research Fellowship for Young
Scientists.
The work of TN was supported in part by JSPS Grant-in-Aid for Young Scientists (B) No.\,15K17628.

\appendix

\section{On holographic calculation of $C(1)$ using graviton propagator}\label{ap:Graviton}
In this appendix, we use the expression \eqref{HolC} including the graviton propagator to calculate $C(1)$ for a spherical entangling surface.
First, we reproduce the formula
\begin{align}
  C(n) &=
  \frac{\pi}{4G_Nn}\int d^{d-1}y\, d^{d-1}y'\,
  \sqrt{g(y)}\sqrt{g(y')}
  J(y,y')\,,
\end{align}
where
\begin{align}
  J(y,y') &\equiv
  \frac{\partial X^\mu}{\partial y^i}
  \frac{\partial X^\nu}{\partial y_i}
  G_{\mu\nu\alpha\beta}(X(y),X(y'))
  \frac{\partial X^\alpha}{\partial y'^j}
  \frac{\partial X^\beta}{\partial y'_j}\,.
\end{align}
There is a difficulty related to the boundary term in this formula as commented in \cite{Chang:2013mca} and pointing it out is the purpose of this appendix.

The graviton propagator $G_{\mu\nu\alpha\beta}$ is not known for the backreacted metric with general $n$, while the metric is just $AdS_{d+1}$ for $n=1$ whose graviton propagator $G_{\mu\nu\mu'\nu'}(X,X')$ can be represented as \cite{D'Hoker:1999jc}
\begin{align}
  G_{\mu\nu\mu'\nu'}(X,X') =
  (\partial_\mu\partial_{\mu'} D\, \partial_\nu\partial_{\nu'} D
  +(\mu\leftrightarrow \nu))G(D)+G_{\mu\nu}(X)G_{\mu'\nu'}(X')H(D) +\cdots\,.
\end{align}
The ($\cdots$) terms are gauge-dependent and do not matter when the bulk energy momentum tensor $T_{\mu\nu}$ vanishes at the boundary fast enough, but they would contribute in the current setup because the energy-momentum tensor of the brane does not decay at the boundary. The ($\cdots$) term is too complicated to be taken into account, and we proceed without having them for a moment.

We are going to evaluate the $G(D)$ and $H(D)$ parts. $\partial_{\mu}=\partial/\partial X^\mu$ and $\partial_{\mu'}=\partial/\partial
X'^{\mu'}$ are derivatives with respect to the bulk points $X$ and $X'$.
The two functions $G(D)$ and $H(D)$ are given by
\begin{align}
  G(D) &=
    \tilde{C}_d\left(\frac{2}{D}\right)^d
    F(d,\frac{d+1}{2};d+1;-\frac{2}{D}) \,,\\
  H(D) &=
    -\frac{2(D+1)^2}{d-1}G(D) + 
    \frac{4(d-2)(D+1)}{(d-1)^2}\tilde{C}_d\left(\frac{2}{D}\right)^{d-1}
    F(d-1,\frac{d+1}{2};d+1;-\frac{2}{D})\,,
\end{align}
with a constant
\begin{align}
  \tilde{C}_d=\frac{\Gamma(\frac{d+1}{2})}{(4\pi)^\frac{d+1}{2}d}=\frac{1}{2^dd\,\text{Vol}(\mathbb{S}^d)}\,.
\end{align}
The function $D=D(X,X')$ is the invariant distance  between the two points
$X$ and $X'$,
\begin{align}
  D = \frac{1}{2}\left[
    -(X'_{-1}-X_{-1})^2 + (X'_0-X_0)^2 +(X'_1-X_1)^2+\dots
    +(X'_d-X_d)^2
  \right] \,,
\end{align}
in the Euclidean $AdS_{d+1}$ space realized as an embedding
$
  -X_{-1}^2 + X_0^2 +X_1^2+\dots+X_d^2 = -1
$ in $\mathbb{R}^{1,d+1}$,
with the metric
$  ds^2 = -dX_{-1}^2 + dX_0^2 + dX_1^2+\dots +dX_d^2
$.
An expression of $D$ in
the hyperbolic coordinate
\begin{align}
  ds_{d+1}^2 = \frac{dr^2}{r^2-1} + (r^2-1)d\tau^2 +r^2(du^2+\sinh^2u d\Omega^2_{d-2})\,,
\end{align}
follows from the coordinate transformation
\begin{align}
\begin{aligned}
  X_{-1} = r\cosh u\,, &\qquad&
  X_{i} &= r\sinh u\, \Omega_{i-1} \quad(i=2,\dots,d)\,, \\
  X_{0} = \sqrt{r^2-1}\sin\tau\,, &\qquad&
  X_{1} &= \sqrt{r^2-1}\cos\tau\,.
\end{aligned}
\end{align}
The minimal surface is the horizon $r=1$
of the topological black hole at $\tau=0$, on which the invariant distance $D$ becomes 
\begin{align}
  D(u',\Omega'_i;u,\Omega_i) &= \cosh u \cosh u'- \sinh u\sinh u'\sum_{i=1}^{d-1}{\Omega_i\Omega_i'}-1\,.
\end{align}
The function $J$ is calculated as\footnote{
To calculate $J=J(D)$, it is easier to work in Poincar\'e coordinate
$  ds_{d+1}^2 = (dz^2 + \sum_{i=0}^{d-1}dx_i^2)/z^2
$
related by
\begin{align}
  X_{-1} = \frac{z}{2}+\frac{1+\sum_{i=0}^{d-1}x_i^2}{2z}, &&
  X_{i} = \frac{x_i}{z} \quad(i=0,\dots,d-1), &&
  X_{d} = \frac{z}{2}+\frac{-1+\sum_{i=0}^{d-1}x_i^2}{2z}\,,
\end{align}
because the minimal surface  is mapped to just a plane $x^0=x^1=0$.

}
\footnote{
The ($\cdots$) term would be represented as
\begin{align}
  (\cdots) &=
    2\big(d(D+1)^2-1\big)X(D) + 2D(D+1)(D+2)X'(D) \nonumber \\
    &\quad +2(d+1)D(D+1)(D+2)Y(D) +2D^2(D+2)^2Y'(D) \nonumber \\
    &\quad +2(d-1)^2(D+1)Z(D) + 2(d-1)D(D+2)Z'(D) \,,
\end{align}
with functions $X,Y,Z$ given implicitly in \cite{Chang:2013mca}.
}
\begin{align}
  J =
    2\Big((D+1)^2+d-2\Big)G(D) + (d-1)^2H(D)+\cdots\,,
\end{align}
which is just a function of the invariant distance $D$.
The symmetry of the hyperbolic space $\mathbb{H}^{d-1}$ allows us to move
the two points to $(u',\Omega') =(0,0)$ and $(u,\Omega)=(u,0)$, 
and factor out the integrals over $u'$, $\Omega'$ and $\Omega$:
\begin{align}
\begin{aligned}
  C(1) &=
  \frac{\pi}{4G_N}\int
  du\,du'\,d\Omega_{d-2}\, d\Omega'_{d-2} \sinh^{d-2} u \sinh^{d-2} u'\,
  J(D) \,,\\
  &=\frac{
    \text{Vol}(\mathbb{H}^{d-1})
   }{4G_N}
   \pi \text{Vol}(\mathbb{S}^{d-2})
   \int
  du\sinh^{d-2} u\,
  J(D)\,,
\end{aligned}
\end{align}
where $D = D(u,0;0,0) = \cosh u-1$.
The integration of $G(D)$ and $H(D)$ parts of $J(D)$
can be performed as
\begin{align}
  C(1) = 
   \frac{\text{Vol}(\mathbb{H}^{d-1})}{4G_N}
   \left(
     \frac{d-2}{d} +\ \cdots
   \right)\,.
\end{align}
Compared with the previous result \eqref{holC(1)}, we speculate that the gauge-dependent part contributes $2/d$.
It would be desirable to include the gauge-dependent contribution in order to confirm our conjecture, but we leave it to future investigations.

\section{Comments on the strong sub-additivity of R\'enyi entropies}\label{ap:SSA_RE}
We have given a holographic proof of the R\'enyi entropic inequalities, but they are not related to the strong sub-additivity of entanglement entropy \eqref{SSA}.
In fact, the R\'enyi entropy $S_n$ is nether strong sub-additive nor sub-additive \eqref{SA}. The improved R\'enyi entropy $\tilde{S}_n$ \eqref{improved1} does not satisfy them too.

To achieve the sub-additivity and strong sub-additivity of the R\'enyi entropy, it would be helpful to review how these inequalities are related to information theoretic measures.
In fact, these inequalities follow from properties of the relative entropy $S[\rho|\sigma]\equiv \Tr[\rho(\log\rho-\log\sigma)]$.
Relative entropy is non-negative $S[\rho|\sigma]\ge0$, and equivalently
the mutual information
\begin{align}
  I(A,B) = S_A+S_B-S_{AB} = S[\rho_{AB}|\rho_A\otimes\rho_B]\ge 0
\end{align}
is also non-negative.
These are also equivalent to the sub-additivity of entanglement entropy.
On the other hand, the strong sub-additivity of entanglement entropy is equivalent to the non-negativity of the conditional mutual information
\begin{align}
\begin{aligned}
  I(A,B|C) &\equiv  S_{A|C}+S_{B|C}-S_{AB|C}\,, \\
  &= S_{AC}+S_{BC}-S_{ABC} -S_C\,,
\end{aligned}
\end{align}
where $S_{X|Y}\equiv S_{XY}-S_Y$ is called the conditional entropy.
Relative entropy decreases monotonically under partial trace,
$S[\rho|\sigma]\ge S[\Tr_B\rho|\Tr_B\sigma]$, meaning that the conditional mutual information $I(A,B|C)$ is non-negative
\begin{align}
\begin{aligned}
  I(A,B|C) &= S[\rho_{ABC}|\rho_A\otimes \rho_{BC}] - S[\rho_{AC}|\rho_A\otimes\rho_C]\,,\\
  &= S[\rho_{ABC}|\rho_A\otimes \rho_{BC}] - S[\Tr_B[\rho_{ABC}]|\Tr_B[\rho_A\otimes\rho_{BC}]]\,,\\
  &\ge0\,.
\end{aligned}
\end{align}

Now one way to generalize these inequalities to the R\'enyi entropy is introducing the relative entropy or mutual information for the R\'enyi entropy.
One promising proposal of the relative R\'enyi entropy is \cite{Wilde:2014eda,muller2013quantum} 
\begin{align}
  S_n[\rho|\sigma] \equiv \frac{1}{n-1}
  \log\Tr[(\sigma^{\frac{1-n}{2n}}\rho\sigma^{\frac{1-n}{2n}})^n]\,, \label{RRE}
\end{align}
which reduces to the relative entropy $S[\rho|\sigma]$ in the limit $n\to1$. This generalization of the relative entropy keeps the non-negativity $S_n[\rho|\sigma]\ge0$ and monotonicity $S_n[\rho|\sigma]\ge S_n[\Tr_B\rho|\Tr_B\sigma]$ under partial trace \cite{frank2013monotonicity}.
So we assert that the R\'enyi generalization of the sub-additivity would be
\begin{align}
  I_n(A,B) \equiv S_n[\rho_{AB}|\rho_A\otimes\rho_B] \ge0\,,
\end{align}
and the R\'enyi generalization of the strong sub-additivity would be
\begin{align}
  S_n[\rho_{ABC}|\rho_A\otimes \rho_{BC}] - S_n[\rho_{AC}|\rho_A\otimes\rho_C] \ge0\,.
\end{align}
For entanglement entropy with $n=1$, these inequalities admit a holographic interpretation.
For the R\'enyi entropy for any $n$, however, it is not possible to express the relative R\'enyi entropy $S_n[\rho|\sigma]$ or R\'enyi mutual information $I_n$ as a linear combination of the R\'enyi entropies. So it is not clear how to interpret these R\'enyi-generalized inequalities holographically, even though we have the holographic R\'enyi  entropy formula. The expression of the relative R\'enyi entropy \eqref{RRE} suggests that it can be calculated by the replica method  \cite{Lashkari:2014yva} and it may have an interpretation and proof of these R\'enyi-generalized inequalities in a holographic system.

\bibliographystyle{JHEP}
\bibliography{HRE}

\end{document}